\documentclass{emulateapj}
\slugcomment{Accepted for publication in ApJ}

\begin{document}
\title{Optical afterglows of short Gamma-ray Bursts and GRB 040924}
%\footnote{}

\author{Y. Z. Fan$^{1,2,3}$, Bing Zhang$^{1}$, Shiho Kobayashi$^{4,5}$ and
 Peter M\'{e}sz\'{a}ros$^{4,5}$}
\affil{$^1$ Dept. of Physics, University of Nevada, Las Vegas, NV
89154, USA.\\ 
$^2$ Purple Mountain Observatory, Chinese Academy of
Science, Nanjing 210008, China.\\
$^3$ National Astronomical Observatories, Chinese Academy of
Sciences, Beijing, 100012, China.\\
$^4$ Dept. of Astronomy and Astrophysics, Pennsylvania State
University, 525 Davey Laboratory, University Park, PA 16802, USA.\\ 
$^5$ Dept. of Physics, Pennsylvania State
University, 104 Davey Laboratory, University Park, PA 16802, USA.} 

\begin{abstract}
Short-duration Gamma-ray bursts (GRBs) ($\leq 2{\rm s}$) have remained
a mystery due to the lack of afterglow detection until
recently. The models to interpret short GRBs invoke distinct
progenitor scenarios. Here we present a generic analysis of short GRB
afterglows, and calculate the optical lightcurves of short GRBs
within the framework of different progenitor models. We show that all
these optical afterglows are bright enough to be detected by the
Ultraviolet and Optical Telescope (UVOT) on board the {\em Swift}
observatory, and that different models could be distinguished with a
well-monitored lightcurve. We also model the afterglow data of the
recently discovered short burst GRB 040924. We find that the  
limited data are consistent with a low medium density environment
which is consistent with the pre-concept of the compact-star merger
progenitor model, although the models with a
collapsar progenitor are not ruled out.
\end{abstract}


%%%%%%%%%%%%%%%%%%%%%%%%%%%%%%%%%
\section{Introduction}
In the past several years great advances have been been made in
revealing the nature of Gamma-ray Bursts (GRBs) of relatively long
duration, i.e. $T_{90}>2{\rm s}$ (e.g. M\'{e}sz\'{a}ros 2002; Zhang \&
M\'{e}sz\'{a}ros 2004 for recent reviews). However, another category
of GRBs, i.e. those with short durations (i.e. $T_{90}<2{\rm s}$),
which comprise about 1/3 of the total GRB population, have remained
as mysterious as long GRBs were before 1997. This has been mainly 
due to the lack of afterglow detections for short GRBs, until very
recently. 

The leading progenitor model for short GRBs invokes merger of two
compact objects (e.g. NS-NS merger or BH-NS merger, Eichler et al. 1989;
Paczy\'{n}ski 1991; Narayan et al. 1992;  M\'{e}sz\'{a}ros \& Rees
1992), which has been found suitable to interpret many short GRB
properties (Ruffert et al. 1997; Popham et al. 1999; Perna
\& Belczynski 2002; Rosswog et al. 2003;  Aloy et al. 2004). 
In this scenario, the burst
site is expected to have a large offset from the host galaxy due to
asymetric kicks during the birth of NSs (Bloom et al.
1999; but see Belczynski et al. 2002), so that the number
density of the external medium in the GRB environment is low,
typically $\sim 10^{-2}{\rm cm^{-3}}$. Alternatively, with the
increasing evidence that long GRB progenitors are collapsars,
it has been suggested that short GRBs may also be associated with
collapsars, with either a less energetic jet (i.e. short emerging
model, Zhang, Woosley \& MacFadyen 2003) or a jet composed of many
subjets seen by an off-axis observer looking into one or a few
subjet(s) (subjets model, Yamazaki, Ioka \& Nakamura 2004). If this 
is the case, the environment around the progenitor should be similar 
to that of long GRBs, which is either a constant density medium 
(e.g. Panaitescu \& Kumar 2002; Yost et al. 2003) with
ISM number density $n\sim 1{\rm cm^{-3}}$ or a prestellar wind
(e.g. Chevalier \& Li 2000). Other possibilities for the origin of
short GRBs have been proposed within the cylindrical jet model
(Wang et al. 2004) and the Poynting-flux dominated GRB model
(Lyutikov \& Blandford 2003).
 
Within the standard afterglow model and adopting a typical compact star
merger environment, the forward shock afterglow emission of short GRBs
have been calculated by Panaitescu et al. (2001), Perna \& Belczynski
2002, and Li et al. (2003). Panaitescu et al. (2001) have shown
that the afterglows of short GRBs are faint, and they are likely to be
most easily detected in the X-ray band. Li et al. (2003) considered
possible $e^\pm$ pair loading and evaluated its possible observational
signature. In this work, we present a generic treatment of short GRB
optical afterglows which differs from the previous ones by including
both the forward and the reverse shock emission, a crucial ingredient
for characterizing the early afterglow light curve and the
spectrum. The model is applied to various progenitor models and sample
lightcurves are calculated which are compared against Swift UVOT
sensitivity (\S2). Lately, a short, soft burst GRB 040924 was located
by HETE-2, which led to the discovery of its optical afterglow (Fox \&
Moon 2004). We also apply the model to fit the afterglow data of this
burst (\S3).

\section{The afterglow of short $\gamma-$ray bursts}   

In the standard afterglow model for a fireball interacting with a
constant density medium (e.g., Sari, Piran \& Narayan
1998), for the forward shock (FS) emission, the cooling frequency
$\nu_{\rm c}^{\rm f}$, the typical synchrotron frequency $\nu_{\rm
m}^{\rm f}$ and the maximum spectral flux $F_{\rm \nu,max}^{\rm f}$
read 
\begin{equation}
\nu_{\rm c}^{\rm f}=4.3\times 10^{17}{\rm
Hz}~E_{51}^{-1/2}\epsilon_{\rm B,-2}^{-3/2}n_{-2}^{-1}{t}_{\rm
d}^{-1/2}({2\over 1+z}), 
\end{equation}  
\begin{equation}
\nu_{\rm m}^{\rm f}=3.9\times 10^{11}{\rm
Hz}~E_{51}^{1\over 2}\epsilon_{\rm B,-2}^{1\over 2}\epsilon_{\rm
e,-0.5}^2{t}_{\rm d}^{-{3\over 2}}[{13(p-2)\over 3(p-1)}]^2({2\over 1+z}), 
\end{equation}
\begin{equation}
F_{\rm \nu,max}^{\rm f}=8.3 {\rm \mu Jy}~E_{51}\epsilon_{\rm
B,-2}^{1/2}n_{-2}^{1/2}D_{28.34}^{-2}({1+z\over 2}),
\end{equation}
where $E$ is the isotropic energy of the outflow, $\epsilon_{\rm e}$ and
$\epsilon_{\rm B}$ are the fractions of the shock energy given to the
magnetic field and electron at the shock, respectively, $n$ is the
number density of the external medium, $p\sim2.3$ is the power-law
distribution index of shocked electrons, $D$ is the luminosity
distance, and $z$ is the redshift. Hereafter $t=t_{\rm obs}/(1+z)$ denotes
the observer's time corrected for the cosmological time dilation
effect, and $t_{\rm d}$ is in unit of day.  
The superscript ``f'' (``r'') represent the forward
(reverse) shock emission respectively. Throughout this work, we adpot
the convention $Q_{\rm x}=Q/10^{\rm x}$ using cgs units.
We have normalized the parameters to typical values of short GRBs. 
The above equations apply to an isotropic fireball, or to a jet 
with opening angle $\theta_0$ when the bulk Lorentz factor $\gamma >
1/(\sqrt{3}\theta_0)$, so that $\gamma\approx
8.2E_{51}^{1/8}n_{-2}^{-1/8}{t}_{\rm d}^{-3/8}$ is satisfied.
If sideways expansion is important, for $\gamma\leq
1/(\sqrt{3}\theta_0)$, one has $\gamma=(\sqrt{3}\theta_0)^{-1} (t_{\rm
d}/t_{\rm 0,d})^{-1/2}$, $F_{\rm \nu,max(J_{\rm s})}^{\rm f}=F_{\rm
\nu,max}^{\rm f}(t_{\rm d}/t_{\rm 
0,d})^{\rm -1}$, $\nu_{\rm c(J_{\rm s})}^{\rm f}\approx \nu_{\rm
c}^{\rm f}(t_{\rm 0,d})$ and $\nu_{\rm m(J_{\rm s})}^{\rm f}\approx
\nu_{\rm m}^{\rm f}(t_{\rm 0,d})({t}_{\rm d}/t_{\rm 0,d})^{-2}$
(Rhoads 1999; Sari, Piran \& Halpern 1999). If sideways expansion is
unimportant, 
equations (1-2) still hold and equation (3) should be replaced by
$F_{\rm \nu,max(J)}^{\rm f}\approx F_{\rm \nu,max}^{\rm f}(t_{\rm
d}/t_{\rm 0,d})^{-3/4}$. Here the subsecript $J$ ($J_{\rm s}$)
represnts a jet without (with) significant sideways expansion,
respectively. During the reverse shock crossing process, the bulk LF
of the ejecta is nearly a constant if the reverse shock is
non-relativistic (which is the case for short bursts). We have $F_{\rm
\nu, max}^{\rm f}\propto t^3$, $\nu_{\rm c}^{\rm f}\propto t^{-2}$ and
$\nu^{\rm f}_{\rm m}$ is independent on $t$.  

The time when RS crosses the shell can be
estimated by $t_{\rm \times}=\max[t_{\rm dec}, T_{90,\rm obs}/(1+z)]$
(Kobayashi, Piran \& Sari 1999). The typical
duration of short bursts is $T_{90, \rm obs}\sim 0.2{\rm s}$, which is much
smaller than the deceleration time $t_{\rm dec}$ for the ISM case.
We therefore have a typical thin-shell regime.
the RS is only mildly-relativistic at the shock crossing time (e.g.
Sari \& Piran 1999; Kobayashi 2000). The typical deceleration radius
is defined as $R_{\rm dec}\approx 5.6\times 10^{16}{\rm cm}~E_{51}^{1/3}n_{\rm
-2}^{-1/3}\eta_{2.5}^{-2/3}$ (Rees \& M\'{e}sz\'{a}ros 1992),     
where $\eta\sim 300$ is the initial Lorentz factor (LF) of the outflow. 
At $R_{\rm dec}$, the LF of the outflow drops to
$\gamma_{\times}=\gamma_{\rm dec}\sim 0.6 \eta$, so that
${t}_{\rm dec}\approx R_{\rm dec}/2\gamma_{\rm dec}^2c=30{\rm s}~
E_{51}^{1/3} n_{-2}^{-1/3} \eta_{2.5}^{-8/3}$.

At ${t}_{\rm \times}=t_{\rm dec}$, the LF of the decelerated outflow
relative to the initial one is 
$\gamma_{34,\times}\approx (\eta/\gamma_{\rm \times}+\gamma_{\rm
\times}/\eta)/2=1.13$. 
The typical frequency of the RS emission can be estimated by 
\begin{equation}
\nu_{\rm m}^{\rm r}(t_{\rm \times})={\cal R}_{\rm
B}{(\gamma_{34,\times}-1)^2\over (\gamma_{\times}-1)^2} 
\nu_{\rm m}^{\rm f}(t_{\rm \times})\propto n^{1/2},
\end{equation}
where ${\cal R}_{\rm B}$ is the ratio of the magnetic 
field in the reverse emission region to that in the FS emission
region (Zhang, Kobayashi \& M\'{e}sz\'{a}ros 2003). Since at least for
some bursts (e.g. GRB990123 and GRB021211) 
the RS emission region seems to be more strongly magnetized (e.g. Fan
et al. 2002; Zhang et al. 2003; Kumar \& Panaitescu 2003), here we
adopt two typical values, i.e. ${\cal R}_{\rm B}=5$ and 1, in the
calculations. There are two possibilities for a magnetized flow
(e.g. Fan et al. 2004). The central engine may 
directly eject magnetized shells. Alternatively, the magnetic fields
generated in the internal shock phase may not be dissipated
significantly in a short period of time (e.g. Medvedev et al. 2005),
and they can get amplified again in the RS region. This second effect,
which has been ignored previously, should also play an important role
in calculating the afterglow re-brightening effect in refresh-shocks.  

Following Kobayashi \& Zhang (2003a) and Zhang et al. (2003), we have  
\begin{equation}
\nu_{\rm c}^{\rm r}\approx {\cal R}_{\rm B}^{-3}\nu_{\rm c}^{\rm
f}\propto n^{-1}, 
\end{equation}
\begin{equation}
F_{\rm \nu, max}^{\rm r}(t_{\rm \times})\approx \eta {\cal R}_{\rm B}
F_{\rm \nu, max}^{\rm f}(t_{\rm \times})\propto n^{1/2}. 
\end{equation}
Generally, the R-band flux satisfies $F_{\nu_{\rm R}}(t_\times)\approx
F_{\rm \nu,max}^{\rm r}(t_\times)[\nu_{\rm R}/\nu_{\rm m}^{\rm r}({\rm
t_\times})]^{\rm -(p-1)/2}\propto n^{\rm p+1\over 4}$. In the thin
shell case, the R-band RS flux is $F^r_{\rm \nu_{\rm 
R}}\propto t_{\rm obs}^{\rm 2p}$ for $t_{\rm obs}<(1+z)t_{\rm
\times}$, and is $F^r_{\rm \nu_{\rm R}}\propto t_{\rm obs}^{-2}$ for
$t_{\rm obs}>(1+z)t_{\rm \times}$ (e.g., Sari \& Piran 1999; Kobayashi 
2000). 

If short GRBs are born in a stellar wind (for the collapsar model), for
the FS emission, the cooling frequency $\bar{\nu}_{\rm c}^{\rm f}$,
the typical synchrotron frequency $\bar{\nu}_{\rm m}^{\rm f}$ and the
maximum spectral flux $\bar{F}_{\rm \nu,max}^{\rm f}$ read (Chevalier
\& Li 2000)
\begin{equation}
\bar{\nu}_{\rm c}^{\rm f}=2\times 10^{13}{\rm Hz}~\epsilon_{\rm
B,-2}^{-3/2}E_{51}^{1/2}A_*^{-2}({2\over 1+z})t_{\rm d}^{1/2}, 
\label{Wind1}
\end{equation}
\begin{equation}
\bar{\nu}_{\rm m}^{\rm f}=4.5\times 10^{12}{\rm Hz}~\epsilon_{\rm
e,-0.5}^2\epsilon_{\rm B,-2}^{1/2}E_{51}^{1/2}({2\over 1+z})t_{\rm
d}^{-3/2}, 
\label{Wind2}
\end{equation}
\begin{equation}
\bar{F}_{\rm \nu,max}^{\rm f}\approx 3.8{\rm mJy}~\epsilon_{\rm
B,-2}^{1/2}E_{51}^{1/2}A_*D_{28.34}^{-2}({1+z\over 2})t_{\rm
d}^{-1/2}, 
\label{Wind3}
\end{equation}
where $A_*=(\dot{M}/10^{-5}M_\odot{\rm yr^{-1}})(v_{\rm w}/10^3{\rm
km~s^{-1}})^{-1}$, $\dot{M}$ is the mass loss rate of the progenitor,
and $v_{\rm w}$ is the wind velocity. Here the bar-parameters
represent the wind case.

Equations (\ref{Wind1}-\ref{Wind3}) apply to an isotropic fireball, or
to a jet with opening angle $\theta_0$ when the bulk Lorentz factor
$\gamma > 1/\sqrt{3}\theta_0$, so that $\gamma\approx
3.3E_{51}^{1/4}A_*^{-1/4}{t}_{\rm d}^{-1/4}$ is satisfied.  For
$\gamma\leq 1/\sqrt{3}\theta_0$, if sideways expansion is significant,
the emission properties is similar to the ISM case (Sari et
al. 1999; Chevalier \& Li 2000). If sideways expansion is unimportant,
equations (\ref{Wind1}-\ref{Wind2}) still hold and equation
(\ref{Wind3}) should be replaced by $\bar{F}_{\rm \nu,max(J)}^{\rm
f}\approx \bar{F}_{\rm \nu,max}^{\rm f}(t_{\rm d}/\bar{t}_{\rm
0,d})^{-1/2}$, where $\bar{t}_{\rm 0,d}$ is determined by
$3.3E_{51}^{1/4}A_*^{-1/4}\bar{t}_{\rm
0,d}^{-1/4}=1/\sqrt{3}\theta_0$.

In the wind case, the RS is usually relativistic (e.g., Chevalier \&
Li 2000). The resulting $t_{\times} \sim T_{90}$, and the optical
emission typically drops as $(t/t_\times)^{-3}$ for $t>t_\times$
(Kobayashi \& Zhang 2003b, Kumar \& Panaitescu 2000). For short
bursts, the duration when the reverse shock emission dominates is too
short for any observational interest. In this work, we do not include
the RS emission in the wind models. 
Below we calculate the typical optical-band lightcurves for 
short GRBs within different progenitor models. 

\subsection{Compact star merger model}

The afterglows of short GRBs powered by mergers have been
investigated by Panaitescu et al. (2001) numerically. 
Here we re-calculate the optical
afterglow lightcurve by also taking into account the RS emission.

\begin{figure}
\epsscale{1.0}
\plotone{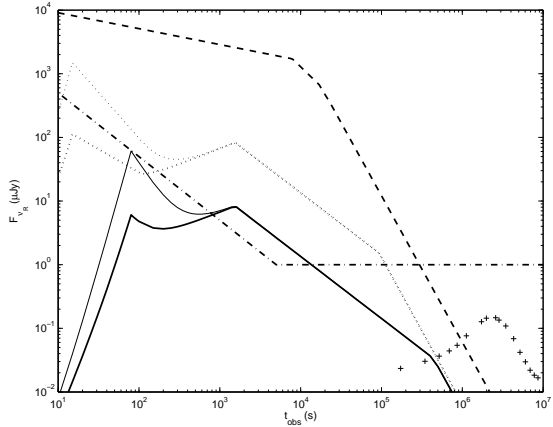}
\caption{The analytical R-band lightcurves of short
GRBs in the compact star merger model and the short emerging
model. The solid lines, dotted lines and the dashed line represent the
compact star merger model in the ISM environment, the short emerging
collapsar model in the ISM environment, and the same model in the wind
environment, respectively. For the first two models (the ISM models),
the reverse shock emission component was calculated for both ${\cal
R}_{\rm B}=5$ (the thin lines) and ${\cal R}_{\rm B}=1$ (the thick
lines). The thick dash-dotted line represents
the sensitivity of UVOT. For $t_{\rm obs}>5000{\rm s}$, the exposure
time of UVOT is assumed to be 1000s, while for $t_{\rm obs}<5000{\rm
s}$, it is assumed to be $t_{\rm
obs}/5$. Following parameters are adopted in the calculations:
$\eta=300$, $\epsilon_{\rm e}=0.3$, $\epsilon_{\rm B}=0.01$,
$p=2.3$, $z=1$, $D=2.2\times 10^{28}{\rm cm}$. In both the compact star
merger model and the short emerging model ISM model, it is assumed that the
outflow is jet-like with an opening angle $\simeq 0.1$ and an
isotropic energy $\simeq 10^{51}{\rm ergs}$. The ISM
number density is taken as $0.01{\rm cm^{-3}}$ and $1{\rm cm^{-3}}$,
respectively. For the short-merging wind model, the density is taken
as $n=3\times 10^{35}R^{-2}{\rm cm^{-3}}$.
For indicative purpose, we also plotted a template 1998bw-like
supernova R-band lightcurve at redshift $z=1$ (the line of plus
signs). }
\label{Light} 
\end{figure}

The lightcurves for this model are plotted as solid lines in Figure
\ref{Light}. At the deceleration time [$\sim
40(1+z){\rm s}$ after the burst trigger], the RS emission reaches 
its peak, and the R band brightness is 20 mag for ${\cal R}_B=5$ 
(thin solid line) and
$z=1$. Swift UVOT has a sensitivity of 24 mag during 1000s exposure
time. Scaling down with time, the sensitivity should be 19 mag for 10s
exposure. Unless the event is much closer or $R_{\rm B}$ is larger,
the RS emission is likely to be below the UVOT sensitivity.  
The FS emission is quite similar to the numerical calculation of
Panaitescu, et al. (2001). Because of a lower $n$ and a
smaller $E$, the R-band afterglow is much dimmer than that of typical
long GRBs, but it is still detectable by the UVOT for at least a few hours.
In the compact star merger scenario,  the collimatation of the
outflow is quite uncertain. Here we adopt $\theta_0\sim 0.1$ as
suggested in numerical simulations (e.g. Aloy et al. 2004). As shown
in Figure \ref{Light}, the lightcurve break occurs too 
late to be detected with the current telescope sensitivity.

\subsection{Short emerging model}

In the ``short emerging model'' (Zhang, Woosley \& MacFadyen 2003),
physical parameters (including the medium density $n$ and the jet opening
angle $\theta_{\rm 0}\simeq 0.1$) are generally similar to the familiar
long GRBs, except that the isotropic energy is smaller. This model has 
received support from a recent comparison study of the spectral properties 
of long and short GRBs (Ghirlanda, Ghisellini \& Celotti 2004).
The R-band lightcurves of this model are plotted as dotted lines in
Fig.\ref{Light} for ISM case, where the thin and thick lines are for ${\cal 
R}_{\rm B}=5$ and ${\cal R}_{\rm B}=1$, respectively. Compared with
the compact star merger model, thanks to a larger $n$ 
($F_{\rm \nu,max}^{\rm f}\propto n^{1/2}$ and $\nu_{\rm m}^{\rm
f}\propto n^0$ for $\gamma\geq (\sqrt{3}\theta_0)^{-1}$), the RS peak
flux is above the UVOT threshold, if ${\cal R}_{\rm B}$ is somewhat
larger than unity. The
RS emission peaks earlier (due to a smaller deceleration radius) so
that the RS peak may be missed if it is shorter than the slewing
time. In any case, the
$t_{\rm obs}^{-2}$ decaying component can be detected for ${\cal
R}_{\rm B}=5$ for a $z=1$ burst. In the wind case, for standard
parameters (e.g. 
$n=3\times 10^{35}R^{-2}{\rm cm^{-3}}$ or $A_*=1$), 
the resulting R-band lightcurve is very bright (see the thick dashed
line plotted in Fig. \ref{Light}), thanks to a relative denser medium
at $R<5.5\times 10^{17}{\rm cm}$.

\subsection{Subjet model}

\begin{figure}
\epsscale{1.0}
\plotone{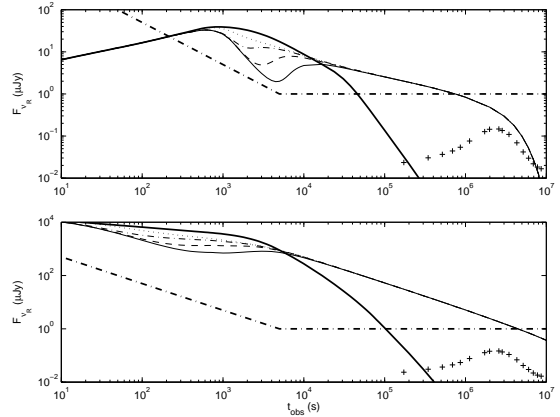}
\caption{The R-band lightcurves of short GRBs for the subjet
model. The typical lightcurve for the short emerging model is also
plotted for comparison. The upper panel is for the ISM case ($n=1{\rm
cm^{-3}}$), and the lower one is for the wind 
case ($n=3.0\times10^{35}R^{-2}{\rm cm^{-3}}$). The thin lines are for
the subjet model. 
The dotted, dash-dotted line, dashed line and solid lines represent
$\Delta \theta=0,~0.01,~0.02,~0.03$ respectively. For clarity, only
the forward shock emission is taken into account. Following parameters
are adopted. For the on-beam subjet, the jet opening angle is
$\theta_{\rm sub}=0.02$, and the isotropic energy is $10^{51}{\rm
ergs}$. For the Gaussian jet, the typical Gaussian angle is
$\theta_{\rm c}=0.08$, the maximum angle 
is 0.3, and angle-dependent energy per solid angle reads
$\epsilon=(10^{53}/4\pi)\exp(-\theta^2/2\theta_{\rm c}^2)$. The
line-of-sight angle is $\theta_{\rm v}=0.26$ from the jet axis. 
The thick solid line
is for the short emerging model calculated with the same code to
calculate the subjet model. The thick
dash-dotted line represents the sensitivity of UVOT. Other parameters
such as $\eta$, $\epsilon_{\rm e}$, $\epsilon_{\rm B}$, $p$ and  $z$
are the same as those adopted to calculate Fig. \ref{Light}.
The supernova bump is also illustrated.}
\label{fig:Num}
\end{figure}

In the subjet model (Yamazaki, Ioka \& Nakamura 2004), GRBs are
conjectured as being powered by many intrinsically similar subjets,
and the number of the subjets are 
distributed with angle as a Gaussian function (Zhang \&
M\'{e}sz\'{a}ros 2002), i.e. $n\propto
\exp[-(\theta/\sqrt{2}\theta_{\rm c})^2]$, with the typical Gaussian
angle $\theta_{\rm c}\simeq 0.1$ (Zhang et al. 2004). If an observer
is far away from the jet axis and by chance is on the beam of one
subjet, one detects a short burst. The
global afterglow emission of this model could be then approximated by
that of a Gaussian structured jet superimposed on a uniform
subjet. 
Here we consider two emission components, one on-beam uniform
less-energetic subjet with an opening angle $\theta_{\rm sub}\approx
0.02$), and another stronger and wider Gaussian structured jet with
typical Gaussian angle $\theta_{\rm c}=0.08$ with the line-of-sight
angle $\theta_{\rm v}\simeq 3\theta_{\rm c}$ off-axis.
Since the Gaussian angular distribution is only of statistical sense in the
subjet model (Yamazaki et al. 2004), around the subjet there could be
a ``void'' where the emissivity is below the Gaussian jet model in
order to counterbalance the emissivity excess at the subjet. Here
we approximate this effect by adopting an annular void region of width
$\Delta\theta$ around the subjet axis (i.e. the emissivity is zero
in the range 
from $\theta_{\rm sub}$ to $\theta_{\rm sub}+\Delta \theta$. In view
of the uncertainties, we calculate the lightcurves for $\Delta
\theta=0,~0.01,~0.02,~0.03$, respectively. Following Yamazaki et
al. (2004) we include a maximum Gaussian jet angle $\theta_{\rm j}
=0.3$ in the calculation.

The afterglow lightcurves of structured jets have been
modeled by many authors (e.g., Wei \& Jin 2003; Kumar \& Granot 2003;
Granot \& Kumar 2003; Panaitescu \& Kumar 2003; Salmonson 2003; Rossi
et al. 2004). 
Here we take the simple method
proposed by Wei \& Jin (2003), in which the sideways expansion of the
jet is ignored (see Kumar \& Granot 2003 for justification) but
the ``equal arriving surface'' effect is taken into account.
The jet evolution is quantified by $\gamma=(3\epsilon/n)^{1/2}(m_{\rm 
p}c^2)^{-1/2}[ct/(1-\mu+1/16\gamma^2)]^{-3/2}$ for the ISM case, and
by $\gamma=(\epsilon/3\times 10^{35}A_*)^{1/2}(m_{\rm
p}c^2)^{-1/2}[ct/(1-\mu+1/8\gamma^2)]^{-1/2}$ for the wind
case\footnote{In the wind case, if we define $X \equiv 
\epsilon/(3\times 10^{35}A_*m_{\rm p}c^3t)$, one has
$\gamma=[X(1-\mu)+\sqrt{X^2(1-\mu)^2+4X}]/2$, and the solutions could
be coasted into a simple form.}. Here
$\epsilon=(10^{53}/4\pi)\exp(-\theta^2/2\theta_{\rm 
c}^2)$ is the energy per unit solid angle of the structured jet,
$\mu=\cos \Theta$, $\Theta$ is the angle between the moving direction
of an emitting unit and the line of
sight. The isotropic energy of the on-beam subjet is taken as
$10^{51}{\rm ergs}$.  
The sideways expansion of the on-beam subjet is also ignored. 
At any emission unit, the standard broken power-law
synchrotron spectrum (e.g. Sari et al. 1998) is adopted with $\delta
F^{\rm f}_{\rm \nu,max}\approx 3\sqrt{3}\Phi_{\rm p}(1+z)\delta N_{\rm
e}m_{\rm e}c^2\sigma_{\rm T} B/\{32 \pi^2eD^2[\gamma(1-\beta
\mu)]^3\}$ (Wijers \& Galama 
1999), where $\Phi_{\rm p}$ is a function of $p$ (for $p\simeq
2.3$, $\Phi_{\rm p}\simeq 0.60$), $B$ is the magnetic field generated
at the shock front. In the ISM case, we take the total number of
electrons swept in the solid angle $d\Omega$ as $\delta N_{\rm
e}=d\Omega R^3 n/3$, where $R$ is the radius of the FS front. In the
wind case, $\delta N_{\rm e}=3.0\times 10^{35}Rd\Omega$ is adopted.  

The model lightcurves for the subjet model are plotted separately in
Fig.\ref{fig:Num}. The upper panel is the ISM case and the lower panel
is the wind
case.  For a comparison, the lightcurve of short emerging model is
also plotted in each model (the thick solid line), which is similar to
the analytical result presented in Figure \ref{Light}.  For the subjet
model, at the early times, the R-band emission is dominated by the
on-beam subjet. As the subjet is decelerated so that the Lorentz
factor is of order $\theta_{\rm sub}$, a very early jet break appears
(see Fig \ref{fig:Num} for detail). 
On the other hand, the energetic Gaussian core component contributes
to the emission steadily, becomes progressively important at later
times, and dominates the afterglow level after thousands of seconds.
Because of the progressively important core contribution, the afterglow
decay in the subjet model is much slower than that in the short
merging model. Notice that the subjet model could be
different from the usual Gaussian jet model in which the angular
energy distribution is smooth (e.g. Kumar \& Granot 2003; Rossi et
al. 2004). The possible existence of the void around the subjet may lead to an
afterglow bump (see Figure \ref{fig:Num}). In fact, if $\Delta \theta$
is 0.1 or larger, the whole jet can be approximated as two
distinct components, i.e. a weak on-beam sub-jet and an off-beam
but more energetic uniform core since the result is insensitive to the
detailed structure in the core.  
The bump can be then understood in terms of the off-beam orphan
afterglow models (e.g. Granot et al. 2002).
In our calculations the initial Lorentz factor across the whole jet is
assumed to be independent on the angle (Yamazaki et al. 2004).

For both the short-emerging model and the subjet model, one may
expect a Type Ib/c supernova component (usually a red bump) showing up
a few weeks after the burst trigger, as has been detected in some long
GRBs. For illustrative purpose, we plot in Fig.\ref{Light} and
Fig. \ref{fig:Num} a template 1998bw-like supernova lightcurve at
$z=1$. The afterglows of short bursts are typically fainter than those
of the long ones, so the supernova signature should be easily
distinguishable, especially for the short emerging model. 
For the subjet model, the contamination of the core component may
make the identification of the SN component more difficult. In any
case, if a flattening or bump is detected within weeks for a short GRB
afterglow, it would argue against the compact star merger model.

\section{GRB 040924}

GRB 040924 triggered the High Energy Transient Explorer 2 (HETE-2) on
2004 September 24 at 11:52:11 UT (Fenimore et al. 2004). The burst
lasted $T_{50}\sim 1.2{\rm s}$, and the energy fluence was ${\cal
F}_{\gamma}\sim 7.7\times 10^{-6}{\rm ergs ~cm^{-2}}$ (Fenimore et
al. 2004; Golenetskii et al. 2004). The ratio of the fluence in the
7-30 keV band and in the 30-400 keV band is about 0.6, so that the
burst is classified as an X-ray rich GRB. The burst redshift was
identified as $z=0.859$ (Wiersema et al. 2004). 
The prompt localization of GRB 040924 by HETE-2
allowed follow-up observations of its afterglow at early times (Fox \&
Moon 2004; Li et al. 2004). Fox (2004) detected an optical transient
$\sim 16$ minutes after the trigger at the level of $m_{\rm R}\simeq
18.0$mag. At the same position, Li et al. (2004) detected an optical
transient $\sim 26$ and $\sim 63$ minutes after the trigger at the
level of $m_{\rm R}\simeq 18.3$mag and $19.2$mag, respectively. Later
detections in K-band and R-band have been reported by many groups
(Terada \& Akiyama 2004; Terada, Akiyama \& Kawai 2004; Hu et
al. 2004;  Fynbo et 
al. 2004; Khamitov et al. 2004a, b, c). The radio observation provides
an upper limit of 0.12mJy at $\sim 15$ hours after the burst (van der
Horst 2004). Below we will compare the available data with the models,
aiming at constraining the burst environment and the possible
progenitor. 

\begin{figure}
\epsscale{1.0}
\plotone{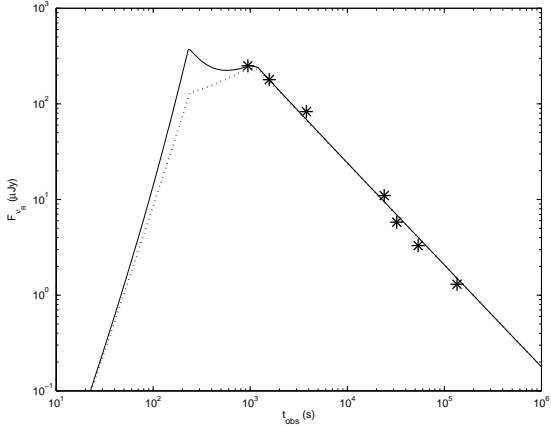}
\caption{Modelling the R-band afterglow data of GRB
040924. The identified burst redshift is $z=0.859$, and the total
fluence is ${\cal F}_{\gamma}=7.7\times 10^{-6}{\rm ergs}$ (Wiersema
et al. 2004). This gives $E_\gamma \simeq 1.5 \times 10^{52} {\rm
erg}$ assuming isotropic emission.
The data (marked by asterisk) are 
taken from Fox (2004), Li et al. (2004), Hu et al. (2004) and
Kahamitov et al. (2004a, b, c). The solid line and dotted line
are the theoretical afterglow lightcurves of a
slow cooling fireball (or a jet with wide opening angle)
expanding into a low density ISM.  
The parameters are $E=3\times 10^{52}{\rm ergs}$, $f_\gamma=2$,
$\epsilon_{\rm e}=0.1$, $\epsilon_{\rm B}=0.004$, $n=0.01{\rm
cm^{-3}}$ and $p=2.42$.
The solid and dotted lines are for ${\cal R}_{\rm
B}=3$ and ${\cal R}_{\rm B}=1$, respectively.} 
\label{Mod}
\end{figure}

\subsection{ISM case}
The constraint $F^{\rm f}_{\rm \nu,max}\geq 250{\rm \mu Jy}$ results in
$f_\gamma{\cal F}_{\gamma,-5.1}\epsilon_{\rm B,-2}^{1/2}n_{-2}^{1/2}\geq
1.3$, where $f_\gamma\geq 1$ is the ratio of the afterglow energy to the
$\gamma-$ray energy. With $z=0.859$ and taking
$f_\gamma=2$, we can estimate $E\simeq 3\times 10^{52}{\rm ergs}$
within the standard cosmology. At the time
$t_{\rm R}\leq 945{\rm s}$, the typical frequency  
of the FS emission crosses the observer frequency
(R-band, $\nu_{\rm obs}=4.6\times 10^{14}{\rm Hz}$). This results in
$0.12[{3(p-1)\over 13(p-2)}]^2(t_{\rm 
R}/945 ~{\rm s})^{3/2}=E_{52.5}^{1/2}\epsilon_{\rm
B,-2}^{1/2}\epsilon_{\rm e,-0.5}^2(1+z)^{1/2}$. We then have the
following constraints 
\begin{eqnarray}
&\epsilon_{\rm e}\leq 0.1[{3(p-1)\over 13(p-2)}]({t_{\rm R}\over
945})^{3/4}E_{52.5}^{-1/4}(f_\gamma/2)^{1\over 2}{\cal F}_{\rm
\gamma,-5.1}^{1\over 2}n_{-2}^{1\over 4},\nonumber\\ &\epsilon_{\rm
B}\geq 4\times 10^{-3} (f_\gamma/2)^{-2}{\cal F_{\rm
\gamma,-5.1}}^{-2}n_{-2}^{-1}.\nonumber
\end{eqnarray}
The observed temporal decay slope is $\alpha_{\rm obs} \simeq -1.07$,
which gives $p=2.42$ in the standard afterglow model. The resulting
spectral index $\beta \simeq -0.71$ matches the observation
$\beta_{\rm obs}=0.61\pm  0.08$ (Silvey et al. 2004). Assuming $t_{\rm
R}\approx 945{\rm s}$ and $F^{\rm 
f}_{\rm \nu,max}= 250{\rm \mu Jy}$ and $n=0.01{\rm cm^{-3}}$, one gets
$\epsilon_{\rm
e}\approx 0.1$ and $\epsilon_{\rm B}\approx 0.004$. The values
of the parameters $\epsilon_e$ and $\epsilon_B$ fall into the regime 
inferred from
afterglow modeling of long bursts (Panaitescu \& Kumar 2002; Yost et
al. 2003). We note that if we take $n\sim 1{\rm cm^{-3}}$,
$\epsilon_{\rm B}\sim 10^{-5}$ is obtained. If the shock 
parameters are more or less universal, our modeling suggests that a
low density ISM model is favored, which is consistent with the
pre-concept of the merger
model. In Fig.\ref{Mod}, we use our model lightcurves to fit the data. 

With parameters derived, $\nu_{\rm c}^{\rm f}$ is above 
the optical energy band in all the observer time, which is consistent
with the observation (Silvey et al. 2004).

\subsection{Wind case}
In the wind case, for $\beta_{\rm obs}\simeq 0.61 \pm 0.08$. With the temporal
index $\alpha_{\rm obs}\simeq -1.07$, $\bar{\nu}_{\rm m}^{\rm f}<\nu_{\rm
obs}<\bar{\nu}_{\rm c}^{\rm f}$ should be satisfied (e.g. Chevalier
\& Li 2000; see also the Tab. 1 of Zhang \& M\'{e}sz\'{a}ros 2004 for
a summary).

At $945$s, the constraints of $\bar{\nu}_{\rm c}^{\rm f}>\nu_{\rm
obs}$, $\bar{\nu}_{\rm m}^{\rm f} \leq \nu_{\rm obs}$ and
$(\bar \nu_{\rm m}^{\rm f} / \nu_{\rm obs})^{\rm (p-1)/2} \bar{F}_{\rm 
\nu,max}=250{\rm \mu Jy}$ yield
\begin{eqnarray}
&A_*<0.14(f_\gamma/2)^{1/4}\epsilon_{\rm B,-2}^{-3/4},\label{Wind4}\\
&\epsilon_{\rm e,-0.5}=0.09g^{1/2}(f_\gamma/2)^{-1/4}\epsilon_{\rm
B,-2}^{-1/4},\label{Wind5}\\ 
&A_*=5.9\times 10^{-4}g^{\rm -(p-1)/2}\epsilon_{\rm
B,-2}^{-1/2}(f_\gamma/2)^{-1/2}.\label{Wind6} 
\end{eqnarray}
By taking
$\epsilon_{\rm B}\sim 10^{-3}$ and $f_\gamma=2$, we have
$\epsilon_{\rm e}\sim 0.05g^{1/2}$ 
and $A_*\sim 1.8\times 10^{-3}g^{\rm -(p-1)}<0.8$, where we have
defined $g=\bar \nu_{\rm m}^{\rm f} / \nu_{\rm obs}$. Therefore,
unless $\epsilon_{\rm e}$ is much smaller than the typical value
$0.1$, we get a very weak stellar wind $A_*\sim 10^{-3}$. A second
problem of the wind 
model comes from the temporal index. For $\beta_{\rm obs}\simeq 0.61$
(Silvey et al. 2004), we have $p\simeq 
2.22$, which in turn results in $\alpha\simeq -1.4$. This is
significantly steeper than $\alpha_{\rm obs}$. We thus suggest that
the wind model is less favored.

In summary, we suggest that the circumburst medium is preferably a
constant density ISM. If we believe that the shock parameters does not
vary significantly among bursts, the inferred $n$ is significantly
lower than that of the typical ISM, which coincides with the
pre-concept of the compact objects meger
model. No definite jet break is detected, so we do not know
the geometrically corrected $\gamma-$ray enengy. If GRB 040924 is
indeed powered by a merger event, no associated Ib/Ic supernova
signature (typically a red lightcurve bump with flux 1$\mu$Jy at
$z\sim 1$) is expected in a few weeks after the burst. The negative
detection of the supernova signature at the time when this work is
completed (two months after the burst trigger) is also consistent with
the compact star merger model.
    
\section{Summary \& Discussion}

We have modeled the typical optical afterglow lightcurves for short
bursts within the context of the leading progenitor models. Both the 
forward and reverse shock emission components are considered. 
With typical parameters, the early afterglows should be detectable 
by the Swift UVOT, and a well-monitored lightcurve can help
to identify the progenitors of short bursts.

The optical afterglow data collected so far for the recent bright
short burst GRB 040924 can be modeled well with an isotropic fireball 
expanding into a low density medium with $n\sim 10^{-2}{\rm cm^{-3}}$.
The wind model is found to be less favored. The resulting  parameters
are consistent with the pre-concept of the compact star merger model. 
Other models such as a collapsar progenitor with low-density
environment, however, cannot be ruled out at this stage.
In principle, if GRB 040924 came from a collapsar, a lightcurve
flattening is expected within weeks resulting from either the
supernova component or the central core component for the subjet
model. The non-detection of such a feature so far presents a further
constraint on the collapsar model.

GRB 040924 is a relatively soft event. It may not be a good
representative of the traditional short-hard bursts. Swift will
locate more short-hard bursts, and our analysis could be directly
utilized to discuss their nature.
 
\acknowledgments YZF thanks D. M. Wei for helpful comments. We also
thank the anonymous referee for helpful suggestions. This work
is supported by NASA NNG04GD51G (for BZ), Eberly Research Funds of
Penn State and by the Center for Gravitational Wave Physics under
grants PHY-01-14375 (for SK), NASA AST 0098416 and NASA NAG5-13286
(for PM), and a NASA Swift GI (Cycle 1) program (for BZ, SK and PM).


\begin{thebibliography}{99}
\bibitem[]{} Aloy, M. A., Janka, H. T., \&  M\"{u}ller, E. 2004, A\&A,
submitted (astro-ph/0408291) 
\bibitem[]{} Belczynski, K., Bulik, T., \& Kalogera, V. 2002, 571, L147
\bibitem[]{} Bloom, J. S., Sigurdsson, S., \& Pols, O. R. 1999, MNRAS,
305, 763\bibitem[]{} Eichler, D., Livio, M., Piran, T., \& Schramm,
D. N. 1989, Nature, 340, 126 
\bibitem[]{} Chevalier, R. A., \& Li, Z. Y. 2000, ApJ, 536, 195
\bibitem[]{} Fan, Y. Z., Dai, Z. G., Huang, Y. F., \& Lu, T. 2002,
Chin. J. Astron. Astrophys. 2, 449 
\bibitem[]{} Fan, Y. Z., Wei, D. M., \& Wang, C. F. 2004, A\&A, 424, 477
\bibitem[]{} Fenimore, E. E., Ricker, G., Atteia, J-L., Kawai, N.,
Lamb, D., \&  Woosley, S. 2004, GCN Circ. 2735
(http://gcn.gsfc.nasa.gov/gcn/gcn3/2735.gcn3) 
\bibitem[]{} Fox, D. B. 2004, GCN Circ. 2741 (http://gcn.gsfc.nasa.gov
/gcn/gcn3/2741.gcn3)
\bibitem[]{} Fox, D. B., \& Moon, D. S. 2004, GCN Circ. 2734 (http://gcn.gsfc.
nasa.gov/gcn/gcn3/2734.gcn3)
\bibitem[]{} Fynbo, J. P. U., Hornstrup, A., Hjorth, J., Jensen,
B. L., \& Andersen, M. I. 2004, GCN Circ. 2747
(http://gcn.gsfc.nasa.gov/gcn/gcn3/2747.gcn3)  
\bibitem[]{} Ghirlanda, G., Ghisellini, G., \& Celotti, A. 2004, A\&A,
422, L55 
\bibitem[]{} Golenetskii, S., Aptekar, R., Mazets, E., Pal'shin, V., 
\& Frederiks, D. 2004, GCN Circ. 2754 (http://gcn.gsfc.nasa.gov/gcn/gcn3

/2754.gcn3)
\bibitem[]{} Granot, J., \& Kumar, P. 2003, ApJ, 591, 1086
\bibitem[]{} Granot, J., Panaitescu, A., Kumar, P., \& Woosley,
S. E. 2002, ApJ, 570, L61  
\bibitem[]{} Hu, J. H., et al. 2004, GCN Circ. 2744 (http://gcn.gsfc.
nasa.gov/gcn/gcn3/2744.gcn3)
\bibitem[]{} Khamitov, I. et al. 2004a, GCN Circ. 2740
(http://gcn.gsfc.nasa.gov/gcn/gcn3/2740.gcn3) 
\bibitem[]{} ------. 2004b, GCN Circ. 2749 (http://gcn.gsfc.nasa.gov/gcn/gcn3
/2749.gcn3)
\bibitem[]{} ------. 2004c, GCN Circ. 2752  (http://gcn.gsfc.nasa.gov/gcn/gcn3
/2752.gcn3)
\bibitem[]{} Kobayashi, S. 2000, ApJ, 545, 807
\bibitem[]{} Kobayashi, S., Piran, T. \& Sari, R. 1999, ApJ, 513, 669
\bibitem[]{} Kobayashi, S., \& Zhang, B. 2003a, ApJ, 582, L75
\bibitem[]{} ------. 2003b, ApJ, 597, 455
\bibitem[]{} Kumar, P. \& Granot, J. 2003, ApJ, 591, 1075
\bibitem[]{} Kumar, P. \& Panaitescu, A. 2000, ApJ, 541, L51
\bibitem[]{} ------. 2003, MNRAS, 346, 905
\bibitem[]{} Li, W., Filippenko, R., Chornock, R., \& Jha, S. 2004,
GCN Circ. 2748 (http://gcn.gsfc.nasa.gov/gcn/gcn3/2748.gcn3) 
\bibitem[]{} Li, Z., Dai, Z. G., \& Lu, T. 2003, MNRAS, 345, 1236
\bibitem[]{} Lyutikov, M., \& Blandford, R. 2003 (astro-ph/0312347)
\bibitem[]{} Medvedev, M. V., et al. 2005, ApJ, 618, L75 
\bibitem[]{} M\'{e}sz\'{a}ros, P. 2002, ARA\&A, 40, 137
\bibitem[]{} M\'{e}sz\'{a}ros, P., \& Rees, M. J. 1992, ApJ,
397, 570
\bibitem[]{} Narayan, R., Paczy\'{n}ski, B., \& Piran, T. 1992, ApJ, 395, L83
\bibitem[]{} Paczy\'{n}ski, B. 1991, AcA, 41, 257
\bibitem[]{} Panaitescu, A., \& Kumar, P. 2002, ApJ, 571, 779
\bibitem[]{} ------. 2003, ApJ, 592, 390
\bibitem[]{} Panaitescu, A., Kumar, P., \&  Narayan, P. 2001, ApJ, 561, L171
\bibitem[]{} Perna, R., \& Belczymski, K. 2002, ApJ, 570, 252
\bibitem[]{} Popham, R., Woosley, S. E., \& Fryer, C. 1999, ApJ, 518, 356
\bibitem[]{} Rees, M. J., \& M\'{e}sz\'{a}ros, P. 1992, MNRAS, 258, 41
\bibitem[]{} Rhoads, J. E. 1999, ApJ, 525, 737
\bibitem[]{} Rossi, E.,  Lazzati,D.,  Salmonson, J. D., \&
Ghisellini, G. 2004, MNRAS, 354, 86
\bibitem[]{} Rosswog, S., Ramirez-Ruiz, E., \& Davies, M. B. 2003,
MNRAS, 345, 1077 
\bibitem[]{} Ruffert, M., et al. 1997, A\&A, 319, 122
\bibitem[]{} Salmonson, J. D. 2003, ApJ, 592, 1002
\bibitem[]{} Sari, R., \& Piran, T. 1999, ApJ, 517, L109
\bibitem[]{} Sari, R.,  Piran, T, \&  Halpern, J. P. 1999, ApJ, 519, L17
\bibitem[]{} Sari, R.,  Piran, T, \&  Narayan, R. 1998, ApJ, 497, L17
\bibitem[]{} Silvey, J. et al. 2004, GCN Circ. 2833 (http://gcn.gsfc.nasa.gov
/gcn/gcn3/2833.gcn3)
\bibitem[]{} Terada, H., \& Akiyama, M. 2004, GCN Circ. 2742
(http://gcn.gsfc.nasa.gov/gcn/gcn3/2742.gcn3) 
\bibitem[]{} Terada, H., Akiyama, M., \& Kawai, N. 2004, GCN
Circ. 2750 (http://gcn.gsfc.nasa.gov/gcn/gcn3/2750.gcn3) 
\bibitem[]{} Wang, X. Y., Cheng, K. S., \& Tam, P. H. 2004, ApJ, in press
\bibitem[]{} Wei, D. M., \& Jin, Z. P. 2003, A\&A, 400, 415
\bibitem[]{} Wijers, R. A. M. J., \& Galama, T. J. 1999, ApJ, 523, 177
\bibitem[]{} Wiersema, K., Starling, R. L. C., Rol, E.,   
Vreeswijk, P.,  Wijers, R. A. M. J. 2004,  GCN Circ. 2800
(http://gcn.gsfc.nasa.gov/gcn/gcn3/ 2800.gcn3) 
\bibitem[]{} van der Horst, A. J., Rol, E., \& Wijers,
R. A. M. J. 2004, GCN Circ. 2746
(http://gcn.gsfc.nasa.gov/gcn/gcn3/2746.gcn3) 
\bibitem[]{} Yamazaki, R., Ioka, K., \& Nakamura, T. 2004, ApJ, 607, L103
\bibitem[]{} Yost, S.,  Harrison,F. A., Sari, R., \&  Frail,
D. A. 2003, ApJ, 597, 459 
\bibitem[]{} Zhang, B., Dai, X., Lloyd-Ronning, N.M. \&
M\'{e}sz\'{a}ros, P. 2004, ApJ, 601, L119
\bibitem[]{} Zhang, B., Kobayshi, S., \&  M\'{e}sz\'{a}ros, P. 2003, ApJ,
595, 950
\bibitem[]{} Zhang, B., \& M\'{e}sz\'{a}ros, P. 2002, ApJ, 571, 876
\bibitem[]{} ------. 2004, Int. J. Mod. Phy. A., 19, 2385
\bibitem[]{} Zhang, W., Woosley, S. E., \& MacFadyen, A. I. 2003,
ApJ, 586, 356 
\end{thebibliography}
\end{document}